\documentclass[structabstract]{aa}  
\usepackage{graphicx}
\usepackage{txfonts}
\usepackage{natbib}
\begin{document}
   \title{Disk and outflow signatures in Orion-KL:
The power of high-resolution thermal infrared spectroscopy\thanks{Based on observations of the ESO program 380.C-0380(A).}}

%   \subtitle{I. Overviewing the $\kappa$-mechanism}

   \author{H.~Beuther
          \inst{1}
          \and
          H.~Linz
           \inst{1}
          \and
          A.~Bik
          \inst{1}
          \and
          M.~Goto
          \inst{1}
          \and
          Th.~Henning
          \inst{1}
%\fnmsep\thanks{Just to show the usage
%          of the elements in the author field}
          }
   \institute{Max-Planck-Institute for Astronomy, K\"onigstuhl 17,
              69117 Heidelberg, Germany\\
              \email{name@mpia.de}
             }

%   \date{Received September 15, 1996; accepted March 16, 1997}

% \abstract{}{}{}{}{} 
% 5 {} token are mandatory
 
%             \abstract{blablabla}

\abstract
{The Orion-KL region contains the closest examples of high-mass
  accretion disk candidates. Studying their properties is an essential
  step in studying high-mass star formation.}
   {Resolving at high spatial and spectral resolution the molecular
     line emission in the immediate environment of the exciting
     sources to infer the physical properties of the associated gas.}
   {We used the CRIRES high-resolution spectrograph mounted on the VLT
     to study the ro-vibrational $^{12}$CO/$^{13}$CO, the
     Pfund\,$\beta$ and H$_2$ emission between 4.59 and 4.72\,$\mu$m
     wavelengths toward the BN~object, the disk candidate
     source n, and a proposed dust density enhancement IRC3.}
   {We detected CO absorption and emission features toward all three
     targets. Toward the BN~object, the data partly confirm the
     results obtained more than 25 years ago by Scoville et al.,
     however, we also identify several new features.  While the
     blue-shifted absorption is likely due to outflowing gas, toward
     the BN~object we detect CO in emission extending in diameter to
     $\sim 3300$\,AU with a velocity structure close to the
     $v_{\rm{lsr}}$. Although at the observational spectral resolution
     limit, the $^{13}$CO line width of that feature increases with
     energy levels, consistent with a disk origin.  If one attributes
     the extended CO emission also to a disk origin, its extent is
     consistent with other massive disk candidates in the literature.
     For source n, we also find the blue-shifted CO absorption likely
     from an outflow.  However, it also exhibits a narrower range of
     redshifted CO absorption and adjacent weak CO emission,
     consistent with infalling motions. We do not spatially resolve
     the emission for source n. For both sources we conduct a
     Boltzmann analysis of the $^{13}$CO absorption features and find
     temperatures between 100 and 160\,K, and H$_2$ column densities
     of the order a few times $10^{23}$\,cm$^{-2}$. The observational
     signatures from IRC3 are very different with only weak absorption
     against a much weaker continuum source. However, the CO emission
     is extended and shows wedge-like position velocity signatures
     consistent with jet-entrainment of molecular gas, potentially
     associated with the Orion-KL outflow system.  We also present and
     discuss the Pfund\,$\beta$ and H$_2$ emission in the region.}
   {This analysis toward the closest high-mass disk candidates
     outlines the power of high spectral and spatial resolution
     mid-infrared spectroscopy to study the gas properties close to
     young massive stars. We will extend qualitatively similar
     studies to larger samples of high-mass young stellar objects to
     constrain the physical properties of the dense innermost gas
     structures in more detail also in a statistical sense.}

\titlerunning{Disk and outflow signatures in Orion-KL}
   \maketitle

\section{Introduction}
\label{intro}

Understanding the physical structure of massive accretion disks is one
of the main unsolved problems in high-mass star formation.  
Although indirect, the main line of arguments for accretion disks
stems from massive molecular outflow observations that identify
collimated and energetic outflows from high-mass young stellar objects
(YSOs, e.g., \citealt{henning2000,beuther2002b,zhang2005,arce2006}).
Collimated jet-like outflow structures are usually attributed to
massive accretion disks and magneto-centrifugal acceleration. Recent 2D
and 3D magneto-hydrodynamical simulations of massive collapsing gas
cores also result in the formation of massive accretion disks
\citep{yorke2002,krumholz2006b,krumholz2009}.  However, it is still
unclear whether such massive disks are similar to their low-mass
counterparts, hence dominated by the central YSO and in Keplerian
rotation, or whether they are maybe self-gravitating non-Keplerian
entities.

While studies at (sub)mm wavelengths are a powerful tool to mainly
study the cold gas and dust components on spatial scales of the order
1000\,AU (e.g., \citealt{cesaroni2006}), such observations are not
that well suited to investigate the inner and warmer components of
massive rotating structures.  In contrast to that, mid-infrared
spectral lines, e.g., ro-vibrationally excited CO emission lines, can
trace these warm gas components.  However, the spectral and/or spatial
resolution was mostly lacking because absorption features were
dominating and hence prohibiting the detection of the accretion disks
in emission. Several recent studies have further demonstrated the
power of high-spectral and high-spatial resolution CO mid-infrared
spectroscopy for disks around low-mass young stellar objects and
Herbig Ae stars (e.g.,
\citealt{goto2006,pontoppidan2008,vanderplas2009}).

To achieve the highest angular and spectral resolution possible, we
observed some of the closest massive disk candidates in Orion at a
distance of 414\,pc \citep{menten2007}~-- Orion-BN (the
Becklin-Neugebauer Object), source n and IRC3~-- in the CO
$v=1-0$ transitions around 4.65\,$\mu$m with the \emph{CRyogenic high
  resolution InfraRed Echelle Spectrograph} (CRIRES,
\citealt{kaufl2004}) at the VLT.  Both objects are well detected at
mid-infrared wavelengths exhibiting various kinds of disk-signatures.

{\bf The Becklin-Neugebauer Object:} Since its detection in the 1960s,
the BN~object is one of the archetypical high-mass YSOs
\citep{becklin1967,henning1990}. \citet{scoville1983} observed the
source in several frequency settings between 2 and 5\,$\mu$m and
detected molecular emission from several CO isotopologues (fundamental
and overtone emission), and they inferred that BN exhibits an
outflow/wind as well as a highly confined region of molecular gas at
high densities and temperatures of $\sim$3500\,K. The estimated
luminosity of the BN~object is $1-2\times 10^4$\,L$_{\odot}$
corresponding to a B0.5 main sequence star \citep{scoville1983}. More
recently, \citet{jiang2005} observed BN in polarized near-infrared
emission, and they also identified signatures caused by an embedded
accretion disk. The BN~object has a high velocity along the line of
sight of $\sim$21\,km\,s$^{-1}$ compared with the cloud velocity of
around 5\,km\,s$^{-1}$ \citep{scoville1993}. This is consistent with
the measured high proper motions of that object (e.g.,
\citealt{plambeck1995}). Whether the BN~object is expelled from the
Trapezium system or during a disintegration of a bound system once
containing source I, source n and the BN~object itself is still a
matter of debate (e.g., \citealt{tan2004,gomez2005,zapata2009}).

{\bf Source n:} Based on a bipolar radio morphology and H$_2$O
maser association, \citet{menten1995} suggested that this source may be
one of the driving sources of the powerful molecular outflows within
Orion-KL. Extended mid-infrared emission was observed perpendicular to
the outflow axis \citep{greenhill2004,shuping2004}, and
\citet{luhman2000} detected CO overtone emission. Both features are
interpreted as likely being due to an accretion disk. Source n is
believed to be in an evolutionary younger stage than the BN~object,
and the luminosity is estimated to be lower as well, of the order
2000\,L$_{\odot}$ \citep{greenhill2004}.

{\bf IRC3:} The source n observations serendipitously covered the
extended infrared source IRC3 (e.g., \citealt{dougados1993}) which we
present here as well. At 3.6\,$\mu$m wavelengths, IRC3 is elongated in
the northeast-southwest direction \citep{dougados1993} and shows
highly polarized near- to mid-infrared emission \citep{minchin1991}.
The observations are consistent with IRC3 being a dust density
enhancement reprocessing light from another source, potentially IRC2
\citep{downes1981,minchin1991,dougados1993}.

\begin{figure}[htb]
\centering
\includegraphics[width=0.49\textwidth]{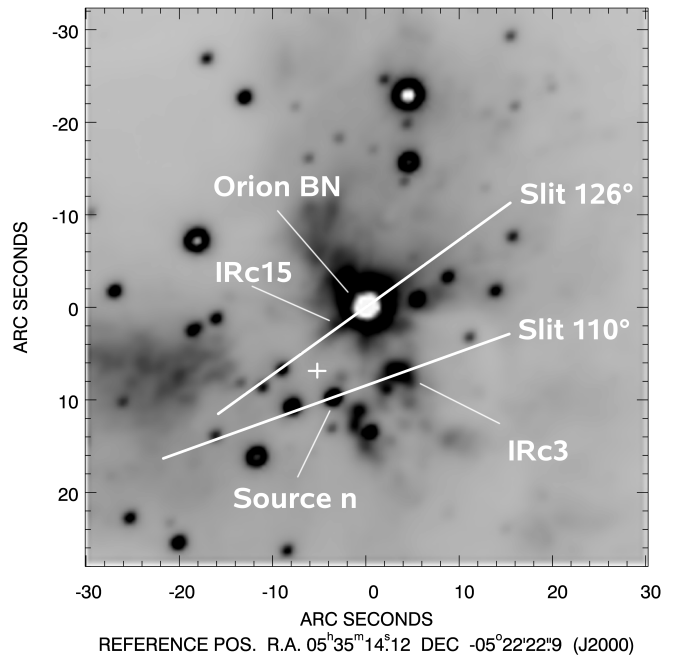}
\caption{Overview image of the region. The grey-scale $K_s$ band image
  is taken from UKIDSS DR5 \citep{lawrence2007}, the white stellar
  peak positions are due to saturation. The slits and important
  sources are marked. The cross denotes the position of IRc2.}
\label{overview}
\end{figure}

Figure \ref{overview} gives an overview of the region marking the
sources discussed in the paper as well as the slit orientations (see
also section \ref{obs}). The nominal absolute positions for the three
sources are listed in Table \ref{positions}.

\begin{table}[htb]
\caption{Source positions (from \citealt{dougados1993})}
\label{positions}
\begin{tabular}{lrr}
\hline
\hline
Source & R.A. & Dec. \\
       & (J2000.0) & (J2000.0) \\
\hline 
BN~object & 05h35m14.12s & -05d22m22.9s \\
Source n  & 05h35m14.35s & -05d22m32.9s \\
IRC3      & 05h35m13.90s & -05d22m30.0s \\
\hline
\hline
\end{tabular}
\end{table} 

\section{Observations}
\label{obs}

We obtained high-resolution spectra between 4.6 and 4.7\,$\mu$m with
CRIRES \citep{kaufl2004}) mounted on UT1 at the VLT on Paranal, Chile.
Two grating settings were selected (12/-1/n,
$\lambda_{\rm{ref}}$=4662.1 and 12/-1/i, $\lambda_{\rm{ref}}$=4676.1)
to observe the spectral interval covering the $^{12}$CO~$v=1-0$
[P(1)--P(5)/R(0)--R(8)] lines without gaps.

One slit covered the BN~object with a position angle of 126 degrees
east of north, whereas the second slit included source n and IRC3 with
a position angle of 110 degrees east of north. For the strong source
BN, only 40\,secs (DIT=2 secs, NDIT=10) were required.  For the second
object with the weaker sources we had a total on-slit integration time
of 20 minutes (DIT=10 secs, NDIT=2).  For the BN observation a slit
width of $0.2"$ was used while for the observation of source n a slit
with of $0.4"$ was selected which correspond to spectral resolving
powers $\lambda /\Delta\lambda$ of 100,000 and 50,000, respectively.
The non-AO mode was applied since no natural guide star is available
in the close environment. The infrared seeing measured from the
spectra was 0.35\arcsec\ during the BN observation and 0.45\arcsec\
for the source n observation. The nod-throw of all the observations
was set to 10\arcsec. To correct for the telluric absorption lines,
attached to every science observation a telluric standard star (HR
1666 with spectral type A3III) was observed.

As extended emission in the CO lines was present in the observations,
we corrected the frames for distortion in order to get a wavelength
solution valid for the whole chip.  Firstly, the chips are slightly
rotated with respect to the slit (ranging from 0.05 degree for chip 3
to 0.45 degree in the case of chip 4). We measured the position of the
brightest object as function of wavelength and calculate from the
displacement in position the rotation angle. Secondly, after the
rotation angle has been corrected, the curvature of the slit is
corrected by measuring the position (central wavelength) of a sky-line
as function of the spatial coordinate. This could be described with a
2nd degree polynomial. Using the IDL routines polywarp and poly\_2d, the
distortion was corrected.

After the distortion correction, the raw files are processed by the
ESO CRIRES pipeline (version 1.10.1) in combination with the Gasgano
software. The data are dark subtracted, flat field corrected as well
as corrected for non-linearity. The wavelength calibration is done
using the telluric emission lines in combination with a HITRAN model
spectrum \citep{rothman2005}. A cross-correlation of the spectra with
the HITRAN spectra showed that the wavelength accuracy of the spectra
is $\pm$0.5\,km\,s$^{-1}$.

Two absorption lines are present in the standard star: the
Pfund\,$\beta$ and the Humphreys\,$\epsilon$ line. Due to the strong
telluric absorption, these lines are not trivial to remove. Therefore,
we first reduced the spectrum without correcting for the intrinsic
absorption lines of the standard star.  In the final reduced spectrum
these lines become eminent as emission lines free from contamination
by the atmosphere.  The Pfund\,$\beta$ line is also seen in our
science object. However, the Humphreys\,$\epsilon$ line is not present
in the science spectrum of BN before division by the standard star.
Therefore, it can be used to correct for the absorption lines of the
standard star.  We used a high resolution Kurucz model spectrum from
an A0V star (http://kurucz.har vard.edu/stars.html) and scaled and
shifted the model spectrum such that the Humphreys\,$\epsilon$ profile
would fit that of the observed Humphreys\,$\epsilon$ profile in the
standard star.  Assuming that the Pfund\,$\beta$ line scales in the
same way as the Humphreys\,$\epsilon$ line does, we divided with the
model spectrum to remove the line contamination.

The spectra were corrected for the earth velocity to the \emph{local
  standard of rest} using rvcorrect in IRAF. The velocity corrections
applied for the two observing dates were -3.8\,km\,s$^{-1}$ for the
BN~data (observed on 21st October 2007) and +43.2\,km\,s$^{-1}$ for
the source n/IRC3 observations (taken on 21st February 2008). The
velocity relative to the local standard of rest $v_{\rm{lsr}}$ of
Orion varies between 2.5 and 9\,km\,s$^{-1}$ (e.g.,
\citealt{comito2005}), and we adopt the approximate value of
+5\,km\,s$^{-1}$.

\section{Results}

\begin{figure*}[htb]
\centering
\includegraphics[width=0.7\textwidth,angle=-90]{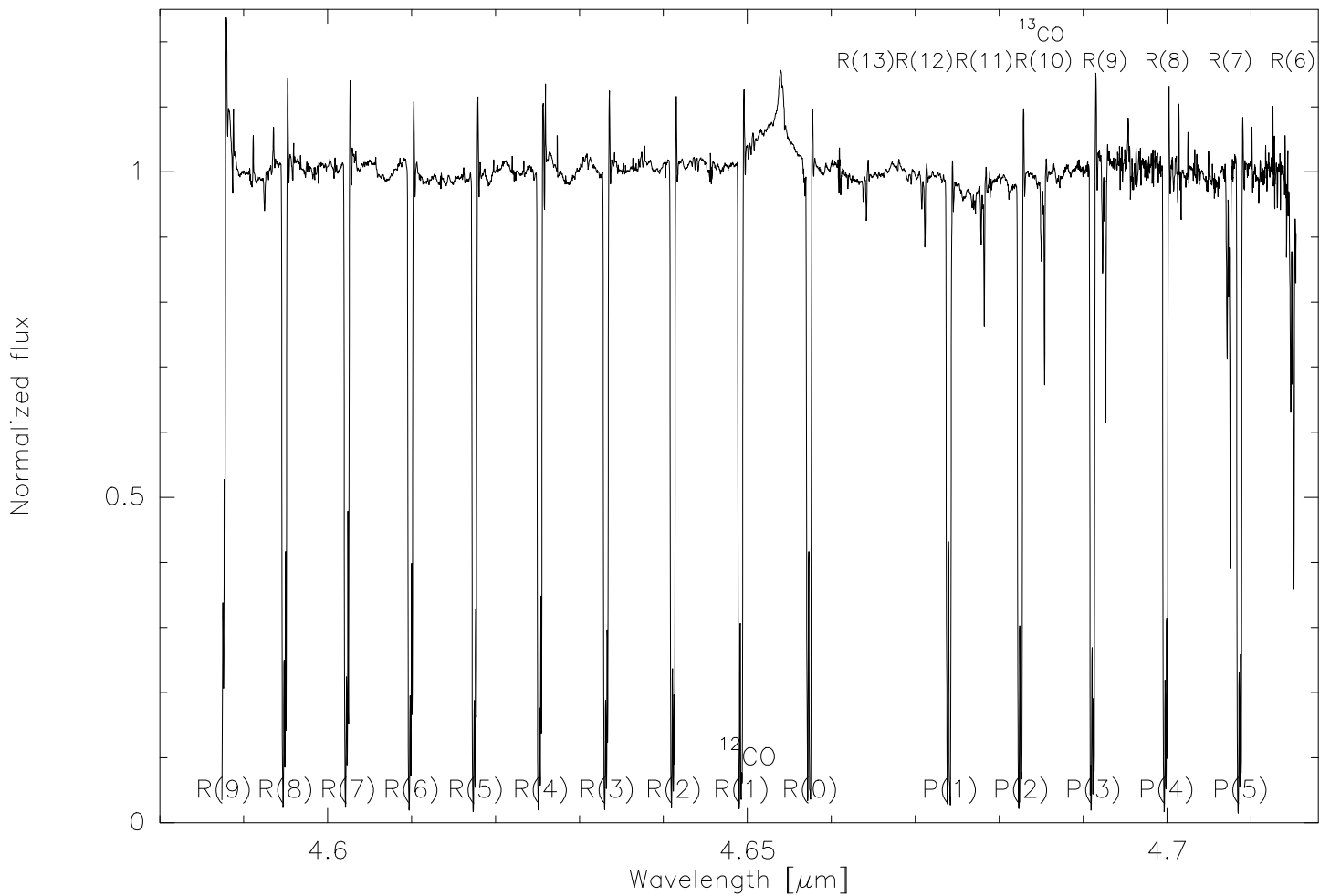}
\caption{CRIRES observations of the R and P CO line series around
  4.65\,$\mu$m. The strong absorption features are the CO lines, and
  the emission line at $\sim$4.654\,$\mu$m is Pfund\,$\beta$.}
\label{bn_all}
\end{figure*}

Toward all three sources we detected the whole suite of
$^{12}$CO\,$v=1-0$ lines present in the spectral window, the
$^{13}$CO\,$v=1-0$ lines from R(6) to R(13) that were not blended by
the $^{12}$CO\,$v=1-0$ lines, as well as the Pfund\,$\beta$ line.
Figure \ref{bn_all} presents the complete spectrum toward the
BN~object, and Table \ref{lines} gives an overview of the covered
lines, their wavelengths $\lambda$ and the lower-level energy state of
the transitions ($E_{\rm{lower}}/k$). In total, this setup covers a
broad range of energy levels extending up to 504\,K.

\begin{table}[htb]
\caption{Observed lines}
\begin{tabular}{lrr}
\hline \hline
Line & $\lambda$ & $E_{\rm{lower}}/k$ \\
     & $\mu$m    & K \\
\hline
$^{12}$CO\\
\hline
R(9) & 4.5876 & 249 \\
R(8) & 4.5950 & 199 \\
R(7) & 4.6024 & 155 \\
R(6) & 4.6090 & 116 \\
R(5) & 4.6176 & 83  \\
R(4) & 4.6254 & 55  \\
R(3) & 4.6333 & 33  \\
R(2) & 4.6412 & 17  \\  
R(1) & 4.6493 & 5.5 \\
R(0) & 4.6575 & 0 \\
P(1) & 4.6742 & 5.5 \\
P(2) & 4.6826 & 17 \\
P(3) & 4.6912 & 33 \\
P(4) & 4.7000 & 55 \\
P(5) & 4.7088 & 83 \\
\hline
$^{13}$CO\\
\hline
% Energies for 13CO taken from 12CO table. Should be the same.
R(13) & 4.6641 & 504 \\	
R(12) & 4.6741 & 432 \\ 
R(11) & 4.6782 & 365 \\ 
R(10) & 4.6853 & 304 \\ 
R(9)  & 4.6926 & 249 \\ 
R(8)  & 4.7000 & 199 \\ 
R(7)  & 4.7075 & 155 \\ 
R(6)  & 4.7150 & 116 \\
\hline
Pfund\,$\beta$ & 4.6538 & $157954^a$\\
H$_2~0-0$\,S9 & 4.6947 & 7198$^a$ \\
\hline \hline
\end{tabular}
~\\
{\footnotesize The temperatures listed for the Pfund\,$\beta$ and H$_2~0-0$\,S9 lines do not indicate gas temperatures since these lines are not excited by collisions within thermal gas, but rather by ionizing photons and within shocked regions, respectively.}
\label{lines}
\end{table}

\subsection{The BN~object}

\begin{figure}[htb]
\centering
\includegraphics[width=0.36\textwidth,angle=-90]{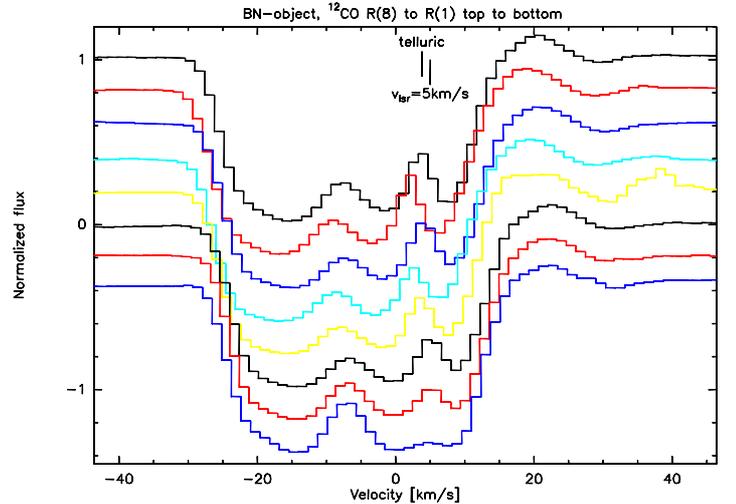}
\caption{The $^{12}$CO R(8) to R(1) lines from top to bottom toward
  the BN~object. Telluric and $v_{\rm{lsr}}$ velocities are marked.
  The feature between 0 and 8 km\,s$^{-1}$ is a telluric artifact.}
\label{bn_rseries}
\end{figure}

\subsubsection{CO absorption and emission}
\label{bn_co}

\begin{figure*}[htb]
\includegraphics[width=0.41\textwidth,angle=-90]{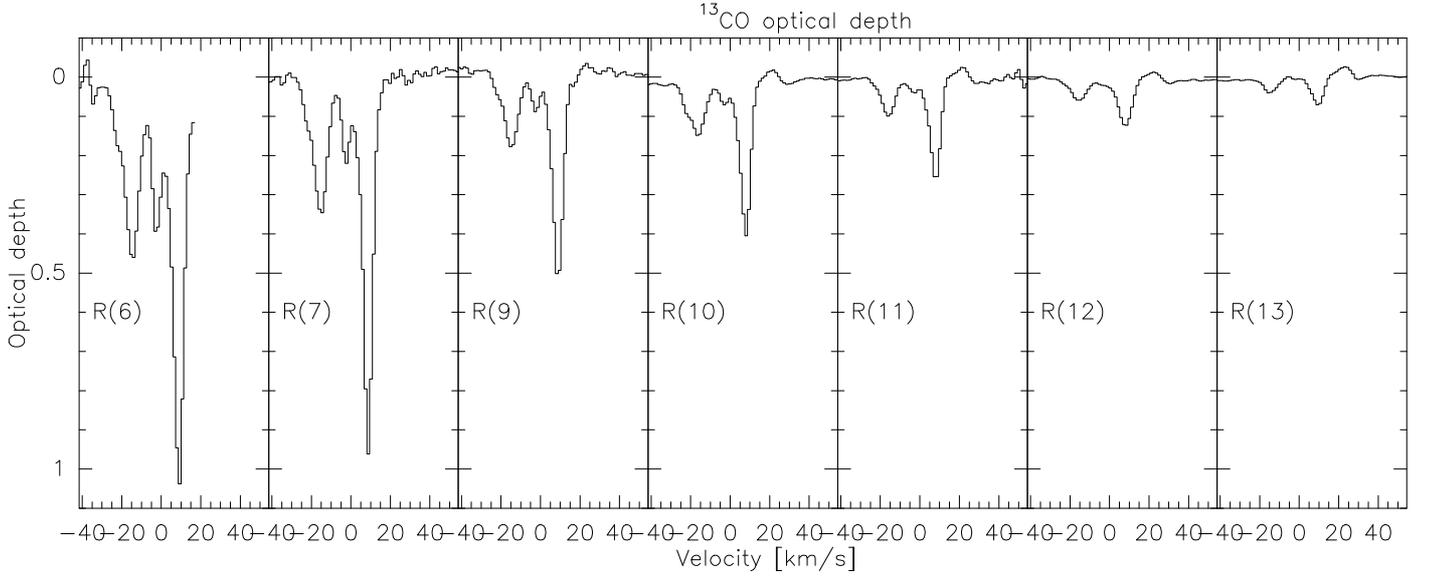}
\caption{$^{13}$CO optical depths toward the BN~object.}
\label{bn_13co}
\end{figure*}

Figure \ref{bn_rseries} presents a zoom compilation of $^{12}$CO data
from the R(1) to R(8) line covering lower energy levels between 5.5
and 199\,K.  If one ignores the telluric line feature at
$\sim$3\,km\,s$^{-1}$ two broad absorption features can be identified
at approximately $-14$ and +8\,km\,s$^{-1}$.  Since all $^{12}$CO
lines are saturated, their peak absorption velocities are unreliable,
and we refer to the $^{13}$CO data (Fig.~\ref{bn_13co}). The peak
velocity of the blue-shifted component is $\sim -15$\,km\,s$^{-1}$
extending from $\sim -28$ to $\sim -8$\,km\,s$^{-1}$. The second
absorption feature has its peak at $\sim 8$\,km\,s$^{-1}$, close to
the $v_{\rm{lsr}}$ of the cloud, and extends from $\sim -8$ to $\sim
18$\,km\,s$^{-1}$. Furthermore, red-shifted from the absorption we
clearly identify a CO emission peaking in $^{12}$CO and $^{13}$CO at
$\sim 20$\,km\,s$^{-1}$ and extending from $\sim$15 to
$\sim$30\,km\,s$^{-1}$. The overall extent of the $^{12}$CO absorption
and emission is from $\sim -30$ to $\sim +30$\,km\,s$^{-1}$. It should
be noted that while the $v_{\rm{lsr}}$ of the different cloud
components for the Orion-KL region vary between approximately 3 and
9\,km\,s$^{-1}$, \citet{scoville1983} inferred that the corresponding
velocity of the BN~object is significantly larger around
21\,km\,s$^{-1}$ (consistent with the different BN ejection scenarios,
e.g., \citealt{tan2004,gomez2005}). While the absorption features stem
from the warmer protostellar envelopes and the surrounding cloud with
usual temperatures of the order 100\,K (see also Boltzmann analysis
below), the ro-vibrational lines in emission can be caused by
different processes. For example, fluorescence via UV photons or
resonance scattering from strong infrared fields can excite these
lines without significantly heating the gas (e.g.,
\citealt{blake2004,ryde2001}). Alternatively, the ro-vibrational lines
could be caused by hotter gas components (see discussion in section
\ref{general}). We note that the critical densities of these lines are
of order $10^{13}$\,cm$^{-3}$ which practically implies that extended
gas components can be hardly responsible for the emission.

Although the blue-shifted part of the spectrum with respect to the
$v_{\rm{lsr}}$ of the molecular cloud is slightly broader than the
red-shifted part, nevertheless we clearly identify red-shifted
absorption as well. To first order, the blue-shifted gas seen in
absorption can be identified with outflowing gas from the region,
whereas the red-shifted features belong to gas infalling in the
direction of the central source. The outflowing gas with a maximum
velocity relative to the $v_{\rm{lsr}}$ of $\sim 35$\,km\,s$^{-1}$ is
consistent with outflow wings often observed at mm wavelengths from
young massive star-forming regions (e.g., \citealt{beuther2002b}). It
should be noted that the even broader outflow wings observed at mm
wavelength toward Orion-KL exceeding $\pm 50$\,km\,s$^{-1}$ (e.g.,
\citealt{chernin1996}) are likely not related to the BN~object but
rather to one or more sources about $10''$ south-east of BN (source I,
source n and/or SMA1, e.g.,
\citealt{greenhill2004,jiang2005,bally2008,beuther2008f}).

How do these general features compare with the data published by
\citet{scoville1983} which were observed during several observing runs
between December 1977 and February 1981, hence about 30 years prior to
our observations. The general CO line structure with two strong
absorption features plus one red-shifted emission peak are largely
the same. Also the overall extent of the emission is quite similar.
Compared to our measured values of $\sim -15$, $\sim +9$ and $\sim
20$\,km\,s$^{-1}$ for the three components, \citet{scoville1983}
report for the corresponding features velocities of $-18$, $+9$ and
$+20$\,km\,s$^{-1}$. While two velocities agree well, the most
blue-shifted absorption peak appears to have shifted a little bit
between the two observations. However, given that their spectral
resolution was more than a factor 2 lower than that of the new CRIRES
data (7 versus 3\,km\,s$^{-1}$), we refrain from further
interpretation of this. Furthermore, \citet{scoville1983} identify two
more absorption features, one at $-3$\,km\,s$^{-1}$ and one at
+30\,km\,s$^{-1}$. Regarding the $-3$\,km\,s$^{-1}$ component we are
not able to infer any changes because that features lies very close to
the telluric emission which obscures any reliable signature there.
However, Fig.~\ref{bn_rseries} shows that we do not detect any
additional absorption feature blue-shifted from the 26.6\,km\,s$^{-1}$
emission. Therefore, the +30\,km\,s$^{-1}$ absorption dip reported by
\citet{scoville1983} was either a transient feature or not significant
with respect to the signal-to-noise ratio.

Since the $^{12}$CO data are so strongly saturated, for the following
analysis we use the corresponding $^{13}$CO data covering the R(6) to
R(13) transitions (Fig.~\ref{bn_all}). Since the absorption depth
$I/I_0$ is related to the optical depth $\tau$ via
$\frac{I}{I_0}=e^{-\tau}$ we can directly estimate the optical depth
of the $^{13}$CO absorption lines shown in Fig.~\ref{bn_13co} if the
lines are spectrally resolved. Since the full width at zero intensity
(FWZI) is of order 20\,km\,s$^{-1}$ (see above), this criterion is
fulfilled with our spectral resolution $\lambda/\Delta\lambda =10^5$
which corresponds to a velocity resolution of 3\,km\,s$^{-1}$. Except
of the lowest $^{13}$CO R(6) line component at $+9$\,km\,s$^{-1}$ all
other observed $^{13}$CO absorption features have optical depths below
1. This allows us to estimate rotational temperatures via Boltzmann
plots from the equivalent line width following the approach outlined
in \citet{scoville1983} also adopting their finite optical depth
corrections. Using their equation (A8), the column density $N_l$ of
the lower-level energy state is

$$ N_l = 1.13 \times 10^{12} \frac{FA}{f_{lu}}\,{\rm{cm}}^{-2}$$

with $F$, $A$ and $f_{ul}$ denoting the optical depth correction
following the appendix in \citet{scoville1983}, the equivalent line
width (in units of cm$^{-1}$) and absorption oscillator strength,
respectively. Because the absorption feature at $\sim$9\,km\,s$^{-1}$
is affected by the telluric line, in the following we only work with
the data from the $\sim -15$\,km\,s$^{-1}$ line. The parameters and
the calculated column densities are listed in Table \ref{values}. In a
Boltzmann distribution, the total column density
$N_{\rm{tot}}(^{13}\rm{CO})$ is related to the temperature and the
lower state column density via:

$$ \ln\left(\frac{N_l}{g_l}\right) = \ln\left(\frac{N_{\rm{tot}}(^{13}\rm{CO})}{Q(T)}\right)-\frac{1}{T_{\rm{rot}}}\frac{E_{\rm{lower}}}{k}$$

with $g_l$ the statistical weight of the lower-level transition and
$Q(T)$ the partition function at the given temperature.

\begin{table}[htb]
\caption{$^{13}$CO values for Boltzmann plots}
\label{values}
\begin{tabular}{lrrrrrr}
\hline
\hline
& $E_{\rm{lower}}/k$ & $A$ & $f_{ul}$ & $F$ & $g_J$ & $N_l(^{13}\rm{CO})$ \\
& K   & km/s    & $\times 10^{-6}$&     &       & $\times 10^{15}$cm$^{-2}$\\
\hline
\multicolumn{2}{l}{BN@-15\,km\,s$^{-1}$}\\
\hline
R(6) & 116 & 3.70 & 6.0 & 1.18 & 13 & 6.0 \\
R(7) & 155 & 3.16 & 5.9 & 1.14 & 15 & 5.0 \\ 
R(9) & 249 & 1.90 & 5.9 & 1.08 & 19 & 2.9 \\
R(10)& 304 & 1.36 & 5.8 & 1.06 & 21 & 2.0 \\
R(11)& 365 & 0.69 & 5.8 & 1.04 & 23 & 1.0 \\
R(12)& 432 & 0.51 & 5.8 & 1.02 & 25 & 0.7 \\
R(13)& 504 & 0.27 & 5.8 & 1.02 & 27 & 0.4 \\
\hline                          
\multicolumn{3}{l}{source n @-7\,km\,s$^{-1}$ comp.}\\
\hline 
R(7) & 155 & 4.1  & 5.9 & 1.14 & 15 & 6.5 \\
R(9) & 249 & 2.7  & 5.9 & 1.08 & 19 & 4.1 \\
R(10)& 304 & 2.0  & 5.8 & 1.08 & 21 & 3.1 \\
R(12)& 432 & 1.3  & 5.8 & 1.06 & 25 & 2.0 \\
R(13)& 504 & 1.5  & 5.8 & 1.05 & 27 & 2.2 \\
\hline                          
\multicolumn{3}{l}{source n @+5\,km\,s$^{-1}$ comp.}\\
\hline 
R(7) & 155 & 4.6  & 5.9 & 1.23 & 15 & 7.8 \\
R(9) & 249 & 2.7  & 5.9 & 1.16 & 19 & 4.4 \\
R(10)& 304 & 1.8  & 5.8 & 1.12 & 21 & 2.8 \\
R(12)& 432 & 0.6  & 5.8 & 1.04 & 25 & 0.9 \\
R(13)& 504 & 0.1  & 5.8 & 1.00 & 27 & 0.1  \\
\hline      
\hline
\end{tabular}
~\\
{\footnotesize The Table entries are explained in the main text.} 
\end{table}

\begin{figure}
  \centering
  \includegraphics[width=0.32\textwidth,angle=-90]{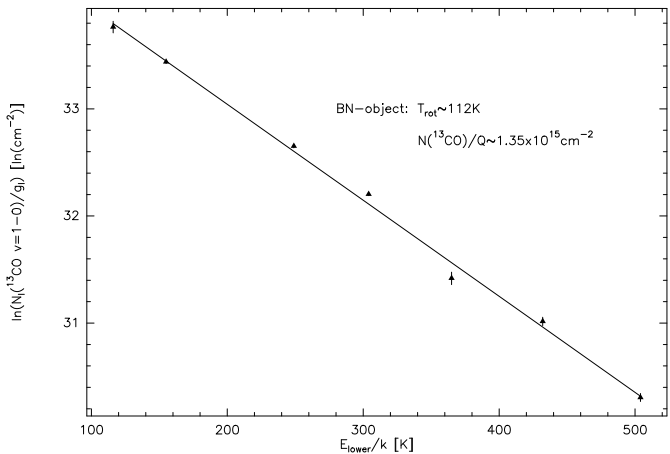}\\
  \caption{Boltzmann plot for the $^{13}$CO v=1-0 lines toward the
    BN~object. The x-axis shows the lower-level energies of the
    transitions and the y-axis presents the natural logarithm of the
    corresponding column densities divided by their statistical
    weights.}
  \label{boltzmann_bn}
\end{figure}

Figure \ref{boltzmann_bn} presents the corresponding Boltzmann plot
where the lower-state column density $N_l$ divided by the statistical
weight $g_J$ is plotted against the lower-state energy
$E_{\rm{lower}}$ divided by $k$. A linear fit to the data gives the
rotational temperature $T_{\rm{rot}}$ and the column density of
$^{13}$CO divided by the partition function $Q(T)$.  For the
$-15$\,km\,s$^{-1}$ $^{13}$CO component the fitted $T_{\rm{rot}}$ is
$\sim 112\pm 20$\,K. \citet{scoville1983} calculated the rotation
temperature for the more blue-shifted absorption component from the
$^{12}$CO$\,v=2-0$ transitions, and they find $\sim$150\,K there,
which is approximately consistent with our new determination. The
other fitted parameter is $N_{\rm{tot}}(^{13}\rm{CO})/Q(T)\approx
1.35\times 10^{15}$\,cm$^{-2}$. For 2-atomic molecules like $^{13}$CO
the partition function can be approximated by $Q(T=112\rm{K})\sim
kT/(hB)\sim 42$ (where $B$ is the rotation constant), and we get a
$^{13}$CO column density of $\sim 5.7\times 10^{16}$\,cm$^{-2}$. Using
furthermore the $^{12}$CO to $^{13}$CO isotopologic ratio of 69 (e.g.,
\citealt{sheffer2007}), we derive a total CO column density of the
$-15$\,km\,s$^{-1}$ component toward Orion-BN of $\approx 3.9\times
10^{18}$\,cm$^{-2}$.  These column density estimates are in excellent
agreement with the results derived by \citet{scoville1983} from the
$^{12}$CO\,$v=2-0$ lines.

How do these values compare to other observations? The BN~object was
detected at 1.3\,mm wavelength by \citet{blake1996} at a 0.15\,Jy
level with a spatial resolution of $1.5''\times 1.0''$.  Following
\citet{plambeck1995}, about 50\,mJy of that flux can be attributed to
circumstellar dust emission.  Assuming optically thin dust emission,
an average dust temperature of 100\,K, a dust opacity index $\beta$ of
2 and a standard gas-to-dust ratio of 100, we can calculate the total
H$_2$ column density
\citep{hildebrand1983,ossenkopf1994,henning1995,beuther2002a,beuther2002erratum,draine2007}.
The derived H$_2$ column density and mass within their $1.5''\times
1.0''$ syntesized beam are then $\sim 1.3\times 10^{24}$\,cm$^{-2}$
and 0.1\,M$_{\odot}$, respectively.  For comparison, using a standard
CO-to-H$_2$ ratio of $8\times 10^{-5}$, the H$_2$ column density
estimated here from the CO IR spectroscopy data is $\sim 4.9\times
10^{22}$\,cm$^{-2}$. While the mm continuum data trace all velocity
components along the line of sight, the near-infrared data trace
selected velocity components. Judging from Figures \ref{bn_rseries}
and \ref{bn_13co}, where we see more than 1 velocity component, the
total column density traced by the CO data is more than a factor 2
higher. Furthermore, the CO absorption only traces the gas in the
foreground of the near-infrared source reducing the traced gas by
another factor 2.  While both approaches are affected by systematics
-- e.g., the mm derived column densities can be wrong by a factor 5
depending on the assumptions on the dust properties and temperatures
-- other reasons are more important for some of the differences. In
particular, the mm continuum emission is sensitive to the cold and
warm dust emission whereas the near-infrared absorption of the
$^{13}$CO transitions useable for our analysis traces mainly the
warmer gas.  Therefore, these data indicate that a large fraction of
the gas is at relatively low temperatures.

\subsubsection{Potential signatures from the BN disk?}

\begin{figure}
  \centering
  \includegraphics[width=0.32\textwidth]{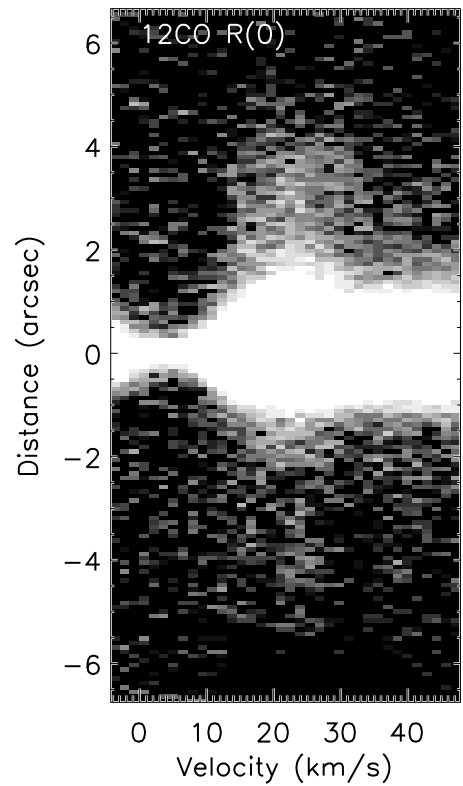}\\
  \caption{Position velocity diagram of the $^{12}$CO (R0) line along
    the slit direction (PA of 126$^{\rm{o}}$ east of north, see
    sec.~\ref{obs}) plotted in logarithmic intensity scale. Around
    offset 0 the figure is dominated by the continuum emission, but
    outside $\sim \pm 2''$ the emission stems from the CO line.
    Positive offsets go in northwest direction.}
  \label{bn_pv}
\end{figure}

While most of the observed emission is spatially unresolved, we find
weak extended $^{12}$CO emission toward the BN~object. Figure
\ref{bn_pv} presents a position-velocity diagram along the slit
direction (PA of 126$^{\rm{o}}$ east of north, see sec.~\ref{obs})
which corresponds to the proposed disk orientation \citep{jiang2005}.
We find CO emission around the rest velocity of the BN~object of $\sim
21$\,km\,s$^{-1}$ extending approximately $\pm 4''$ in both
directions. Since we do not identify any strong velocity dispersion,
this extended emission is unlikely to be due to an outflow, but it may
potentially come from the proposed disk \citep{jiang2005}.  The
measured extent of $\sim 8''$ would then correspond at the given
distance of Orion of 414\,pc to an approximate disk diameter of $\sim
3300$\,AU or a disk radius of 1650\,AU. While such disk size would be
comparably large with respect to typical low-mass disks (e.g., several
reviews in \citealt{reipurth2007}), it is consistent with measured
sizes of rotating structures in other high-mass star-forming regions
(e.g., \citealt{schreyer2002,cesaroni2005,beltran2006b,beuther2008a}).

While we do not identify extended emission is the rarer $^{13}$CO
isotopologue, the CO \& $^{13}$CO emission feature around
20\,km\,s$^{-1}$ exhibits an additional interesting feature in the
$^{13}$CO data (Fig.~\ref{bn_13co}): While the emission feature is at
the edge of the bandpass for the $^{13}$CO R(6) line, it remains
undetected for the R(7) line and only comes up for the lines greater
R(9). This emission feature is also visible in all $^{12}$CO lines
(R(1) to R(8)) shown in Figure \ref{bn_rseries}, but it does not
exhibit any significant shape change there. Likely, the $^{12}$CO
emission is optically thick and traces only an outer envelope that
does not show big variations with excitation temperature. In contrast
to that, the measured Gaussian line width of that component for the
more optically thin $^{13}$CO R(9), R(10), R(11), R(12) and R(13)
lines are 4.3, 5.3, 6.5, 6.4 and 6.6\,km\,s$^{-1}$, respectively.
Although errors on the line width are difficult to quantify because
the emission is at the flank of the strong absorption feature, and
furthermore our spectral resolution is only 3\,km\,s$^{-1}$, the data
are indicative of a line width increases with excitation temperature
of the line that appears to saturate for the highest detectable
transitions R(11) to R(13). While for a centrifugally supported disk
with a central mass of $\sim$10\,M$_{\odot}$ the circular velocity at
3300\,AU is relatively small of order $\sim 1.6$\,km\,s$^{-1}$, the
measured line width increase is consistent with a rotating disk
structure because the inner region with larger rotation velocities
should have higher temperatures caused by the exciting central star.
One has to keep in mind that for purely thermal excitation, rotation
would not cause such an effect because the upper levels -- all in the
3000\,K regime -- do not exhibit big relative excitation differences.
Hence, they do not trace significantly different regions of a disk
then. In contrast to that, for lines that are excited by UV
fluorescence such an effect is possible because the excitation
mechanism via an electronic excited state tends to preserve the level
population of the $v=0$ rotational states (e.g.,
\citealt{brittain2009}). Therefore, in this case the R(9) to R(13)
lines are sensitive to gas temperatures in a relatively broader range
between 150 and 500\,K (Table \ref{lines}). An exact reproduction of
the line width increase would require a detailed disk model, including
its density and temperature structure.  Since we cannot constrain
these parameters well from our data, this is beyond the scope of this
paper. Although the line width increase is below our nominal spectral
resolution element, qualitatively the observations are consistent with
a rotating disk structure.

\subsubsection{The Pfund\,$\beta$ line toward the BN object and H$_2$ emission nearby BN}

\begin{figure}
  \centering
  \includegraphics[width=0.26\textwidth,angle=-90]{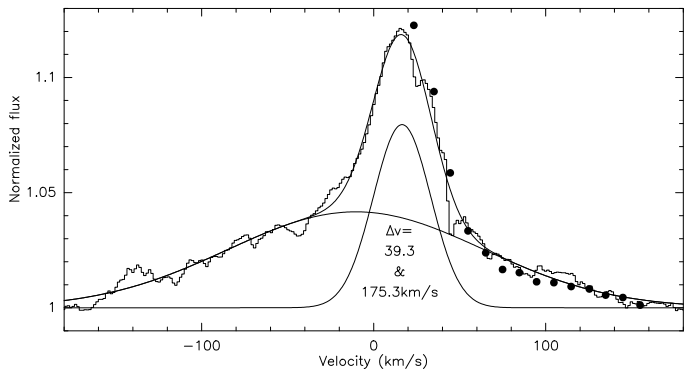}
  \caption{Hydrogen recombination Pfund\,$\beta$ line toward the
    BN~object. The histogram shows the data, and the three other lines
    present a two-component Gaussian fit to the data. Both Gaussians
    are shown separately as well as the combined fit. The FWHM $\Delta
    v$ of both components is given in the figure. The big dots
    reproduce the best fit to the Br$\alpha$ line (their Figure 7
    re-scaled to our normalized Pfund\,$\beta$ spectrum) for an
    optically thin outflow with a velocity law $v \propto r^{-2/3}$ by
    \citet{scoville1983}.}
  \label{bn_pfund}
\end{figure}

Figure \ref{bn_pfund} presents a zoom into the Pfund\,$\beta$ line.
The line shape consists of two components, one central Gaussian
component and broad line wings. It is possible to fit the whole
profile relatively well with a two-component Gaussian fit where the
central component has a FWHM $\Delta v = 39.3$\,km\,s$^{-1}$ and the
broad component has a FWHM of $\Delta v = 175.5$\,km\,s$^{-1}$. The
full width down to zero intensity is approximately 340\,km\,s$^{-1}$
(between -170 and 170\,km\,s$^{-1}$). Close to the peak of the profile
at around 25\,km\,s$^{-1}$, the spectrum exhibits a small dip which is
likely an artifact from the telluric corrections (see section
\ref{obs}). Considering this, the peak of the central Gaussian fit at
$\sim$16.5\,km\,s$^{-1}$ is still consistent with the velocity derived
for the BN~object by \citet{scoville1983} of $\sim$21\,km\,s$^{-1}$.

Figure \ref{bn_pfund} also presents as thick dots the best fit
obtained by \citet{scoville1983} for the Br$\alpha$ line (their Figure
7 re-scaled to our normalized spectrum). Their model consists of a
supersonic, optically thin decelerating outflow with a velocity law of
$v \propto r^{-2/3}$. It is remarkable how well the shape of their
Br$\alpha$ line obtained $\sim$30\,years ago corresponds to the shape
of the newly observed Pfund\,$\beta$ line. Hence, these Pfund\,$\beta$
data are also consistent with their outflow model.

\begin{figure}
  \centering
  \includegraphics[width=0.45\textwidth]{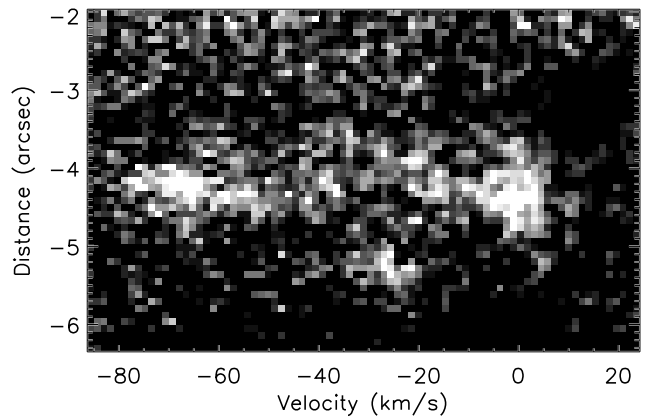}
  \caption{Position velocity digram of the H$_2$ emission $\sim 4.3''$
    south-east of the BN-object associated with the infrared source
    IRC15 \citep{dougados1993}.}
  \label{bn_h2}
\end{figure}

Furthermore, approximately $4''$ south-east of the BN~object we detect
H$_2$ emission from the H$_2~0-0$\,S9 transition with
$E_{\rm{lower}}/k=7198$\,K.  This H$_2$ emission feature is spatially
associated with the corresponding H$_2$ knot in near-infrared H$_2$
images (e.g., \citealt{nissen2007}) as well as with the mid-infrared
source IRc15 reported by \citet{shuping2004}. Figure \ref{bn_h2} shows
a position velocity diagram of this feature. While we do not resolve
any spatial structure of the H$_2$ knot, it shows a very broad
velocity extent going to blue-shifted velocities of $\sim
-80$\,km\,s$^{-1}$, in excess of the CO absorption features measured
toward the BN~object. Since we are mainly interested in the BN~object,
source n and IRC3, we refrain from further analysis of this offset
H$_2$ emission.

\subsection{Source n}

\subsubsection{CO absorption and emission}
\label{n_co}

\begin{figure}
\centering
\includegraphics[width=0.36\textwidth,angle=-90]{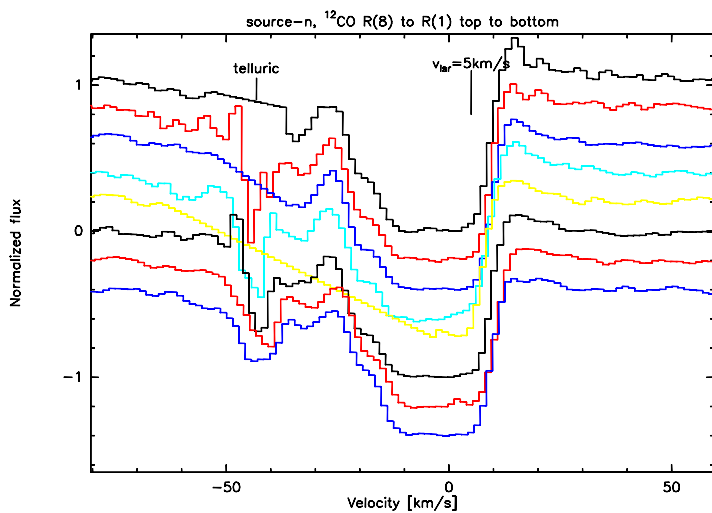}
\caption{The $^{12}$CO R(8) to R(1) lines from top to bottom toward
  source-n. Telluric and $v_{\rm{lsr}}$ velocities are marked. The
  linear slopes and dips around -40\,km\,s$^{-1}$ associated with the
  R(8), R(6) and R(4) lines (for the latter extending from $\sim$-5 to
  $\sim$-65\,km\,s$^{-1}$) are artifacts due to the telluric
  corrections.}
\label{source-n_rseries}
\end{figure}

\begin{figure*}
\centering
\includegraphics[width=0.42\textwidth,angle=-90]{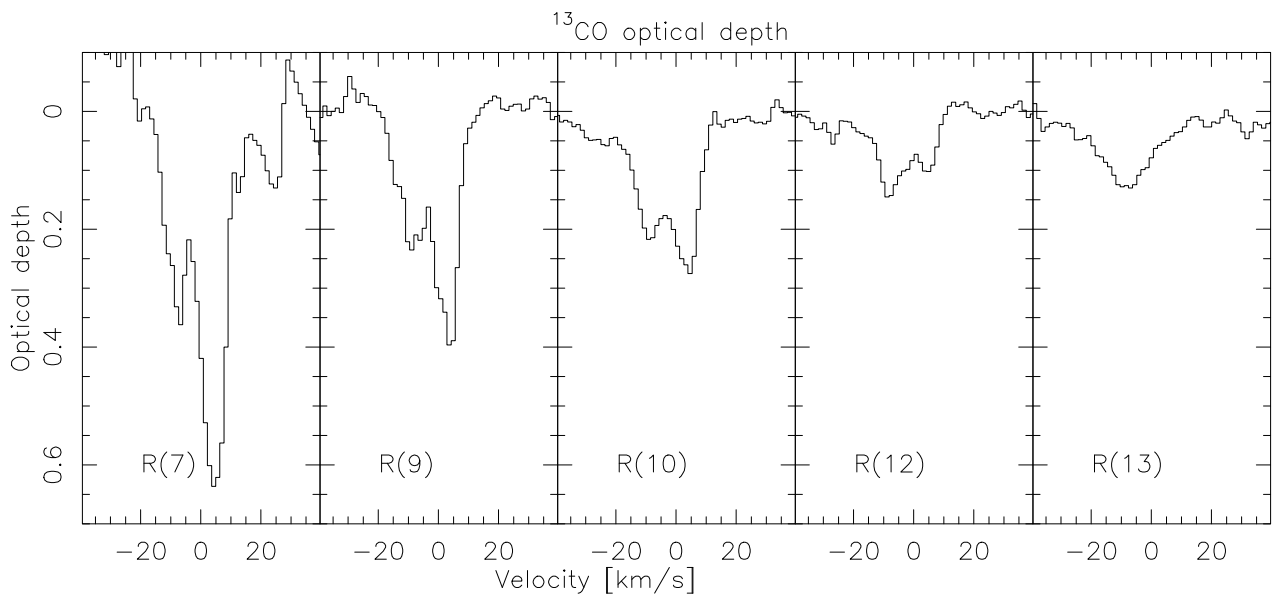}
\caption{$^{13}$CO optical depths toward source-n. The absorption
  feature at $\sim$+25\,km\,s$^{-1}$ of the R(7) line is likely an
  artifact due to telluric lines.}
\label{source-n_13co}
\end{figure*}

Due to the different observing dates, for source n (and IRC3 in the
following section) the telluric lines are almost entirely shifted out
of the CO spectrum. Figures \ref{source-n_rseries} and
\ref{source-n_13co} present the corresponding $^{12}$CO R-series lines
and the detected $^{13}$CO lines with their associated optical depths.
In comparison to the previous BN~data, for source n we only identify
one broad absorption feature in the $^{12}$CO data which is dominated
by blue-shifted outflowing gas. It is interesting to note that the
most blue-shifted absorption does not have a Gaussian shape but rather
a more extended wing-like structure. In the framework of accelerated
winds with distance from the driving star or disk, such a spectral
behavior would be expected. These accelerated winds increase in
velocity with distance from the star. Simultaneously, with increasing
distance from the driving source the density of the surrounding gas
and dust envelope also decreases, lowering the corresponding
absorption depth at those velocities. Hence, in this picture,
higher-velocity gas should show shallower absorption features in the
spectra (e.g., \citealt{lamers1999}).  Furthermore, we also identify a
red-shifted emission component at $\sim$+15\,km\,s$^{-1}$.  However,
compared to the BN~object, where the emission feature can be
identified in all transitions, for source n it is more prominent in
the higher excited lines. The total width of the CO absorption and
emission is $\sim$90\,km\,s$^{-1}$ ranging from approximately
$\sim$-65 to $\sim$+25\,km\,s$^{-1}$ (broader than for BN). The
blue-shifted end is less well determined because of telluric line
contamination.  Nevertheless the blue-shifted outflow part of the
spectrum extends $\sim 70$\,km\,s$^{-1}$ from the assumed
$v_{\rm{lsr}}$ of +5\,km\,s$^{-1}$. This value exceeds that measured
toward BN by about 20\,km\,s$^{-1}$. Although at the edge of the
telluric line contamination, we tentatively identify an additional
discrete absorption feature at $\sim$-35\,km\,s$^{-1}$.

Since the $^{12}$CO line is again always saturated
(Fig.~\ref{source-n_rseries}), for the additional analysis we use the
corresponding $^{13}$CO data (Fig.~\ref{source-n_13co}).  Because of
line-blending and telluric emission, we only measure five $^{13}$CO
lines for source n. The broad, saturated $^{12}$CO absorption feature
is clearly resolved into two components with approximate velocities at
$\sim$-7\,km\,s$^{-1}$ and $\sim$+5\,km\,s$^{-1}$. While the
$\sim$+5\,km\,s$^{-1}$ shows the deeper absorption features for the
lower energy R-lines, it is interesting to note that the single
absorption peak observed for the $^{13}$CO R(13) line is associated
with the $\sim$-7\,km\,s$^{-1}$ component, indicating hotter gas at
more blue-shifted velocities. Following the approach outlined for BN
in section \ref{bn_co}, the derived optical depth of $^{13}$CO is
$\sim$0.6 at its highest.

\begin{figure}
\centering
\includegraphics[width=0.32\textwidth,angle=-90]{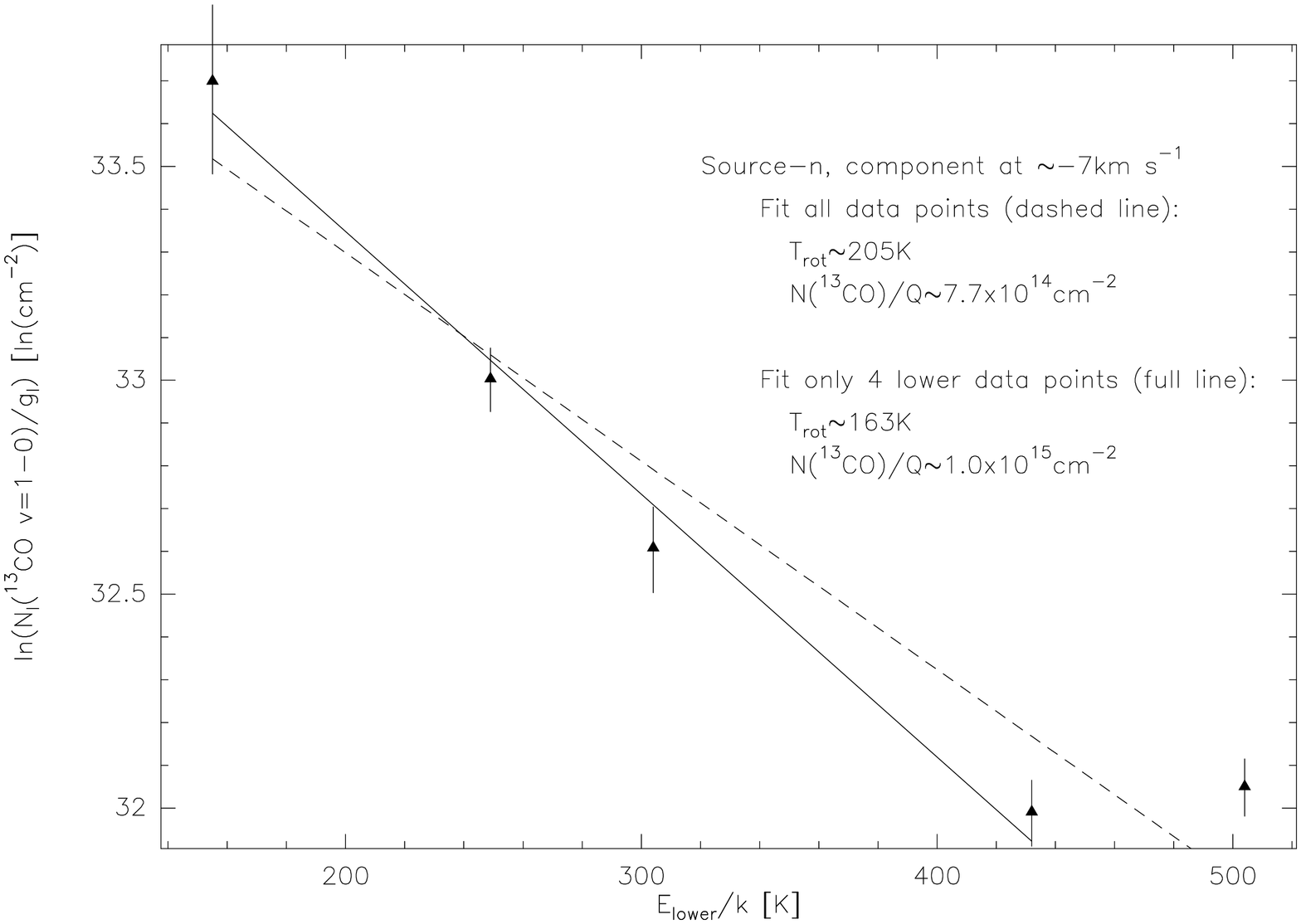}
\includegraphics[width=0.335\textwidth,angle=-90]{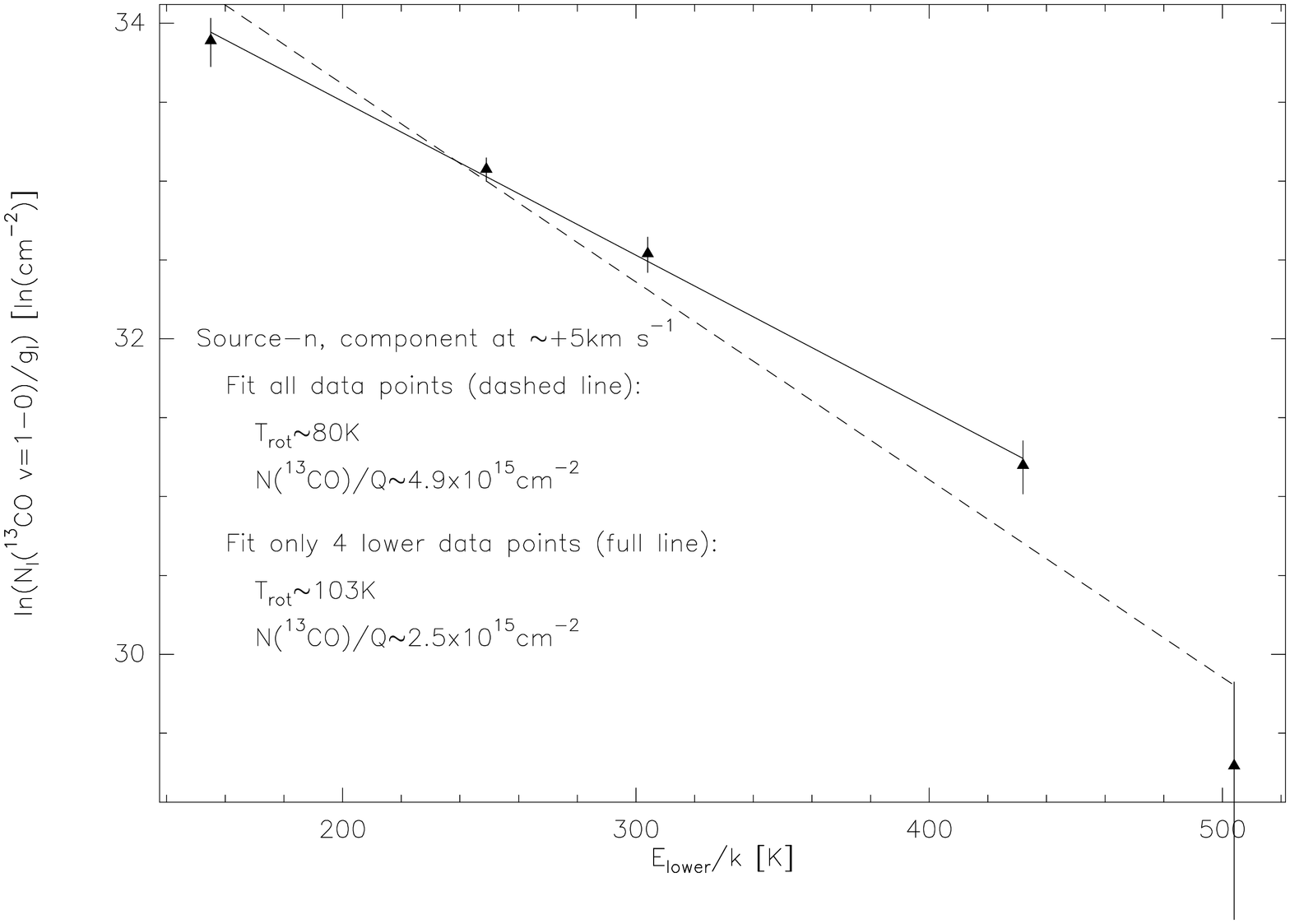}
\caption{Boltzmann plots for the $^{13}$CO v=1-0 lines toward source-n
  for the $\sim$-7\,km\,s$^{-1}$ and the $\sim$+5\,km\,s$^{-1}$
  components in the upper and lower panels, respectively. The x-axis
  shows the lower-level energies of the transitions and the y-axis
  presents the natural logarithm of the corresponding column densities
  divided by their statistical weights. In both panels, the dashed
  line shows a fit to all five data points, whereas the full line
  presents a fit to only the four lower transitions.}
\label{boltzmann_n}
\end{figure}

Since the telluric lines are far away in velocity space (see
Fig.~\ref{source-n_rseries}), for source n we can conduct the
Boltzmann analysis for both absorption components separately.  The
measured equivalent widths $A$ for both components are listed in Table
\ref{values}\footnote{The equivalent width $A$ is measured by
  simultaneous Gaussian fits to both absorption features.}. Again
calculating the $^{13}$CO v=1 column densities (Table \ref{values})
and producing Boltzmann plots (Fig.~\ref{boltzmann_n}), we find that
for both velocity components fits to all 5 data points are less good
than in the case of the BN~object. This is likely due to the fact that
fitting 2 Gaussians to the broad R(13) line, which does not show two
well separated absorption features anymore, may overestimate the
contribution of the $\sim$-7\,km\,s$^{-1}$ component and hence
underestimate the contribution from the $\sim$+5\,km\,s$^{-1}$
component. Therefore, we also fit only the 4 lower-energy transitions
between R(7) and R(12) resulting in better fits
(Fig.~\ref{boltzmann_n}). As expected from the different behavior of
the two absorption components with increasing energy levels, the
derived rotation temperatures using the 4 lower energy lines of the
$\sim$-7\,km\,s$^{-1}$ and $\sim$+5\,km\,s$^{-1}$ are $\sim$163$\pm
20$\,K and $\sim$103$\pm 10$\,K, respectively. As a comparison,
rotational temperatures measured at submm wavelength from CH$_3$OH
emission lines toward source n are around 200\,K \citep{beuther2005a}.
Considering that these measurements are conducted with different
molecules and very different observational techniques, the overall
range of similar temperatures derived at mid-infrared and submm
wavelengths is reassuring for the complementarity of such
multi-wavelengths observations.

Fitting only the 4 lower-energy lines, the column density of $^{13}$CO
divided by the partition function $Q(T)$ results in
$N_{\rm{tot}}(^{13}\rm{CO})/Q(T)\approx 1.0\times 10^{15}$\,cm$^{-2}$
and $N_{\rm{tot}}({13}\rm{CO})/Q(T)\approx 2.5\times
10^{15}$\,cm$^{-2}$ for the $\sim$-7\,km\,s$^{-1}$ and
$\sim$+5\,km\,s$^{-1}$ components, respectively. Approximating again
the partition function by $Q(T)\sim kt/(hB)$ (section \ref{bn_co}) we
have $Q(163\rm{K})\approx 62$ and $Q(103\rm{K})\approx 39$, and using
the $^{12}$CO to $^{13}$CO isotopologic ratio of 69 (e.g.,
\citealt{sheffer2007}), the derived total CO column densities toward
source n are $\approx 4.3\times 10^{18}$\,cm$^{-2}$ and $\approx
6.7\times 10^{18}$\,cm$^{-2}$, of the same order as for the BN object.
With the CO-to-H$_2$ conversion factor of $8\times 10^{-5}$, the
corresponding H$_2$ column densities are $5.3\times
10^{22}$\,cm$^{-2}$ and $8.4\times 10^{22}$\,cm$^{-2}$. Adding these
two column density values, the total H$_2$ column density traced
toward source n by this near-infrared observations is $\sim 1.4\times
10^{23}$\,cm$^{-2}$, more than an order of magnitude below the H$_2$
column densities derived with the Submillimeter Array at 865\,$\mu$m
($\sim 6\times 10^{24}$\,cm$^{-2}$, \citealt{beuther2004g}). Similar
to the BN-case discussed in section \ref{bn_co}, this shows that the
near-infrared data mainly trace the warmer gas whereas the (sub)mm
continuum observations trace warm and cold gas components.

\subsubsection{The Pfund\,$\beta$ line}

\begin{figure}
\centering
\includegraphics[width=0.26\textwidth,angle=-90]{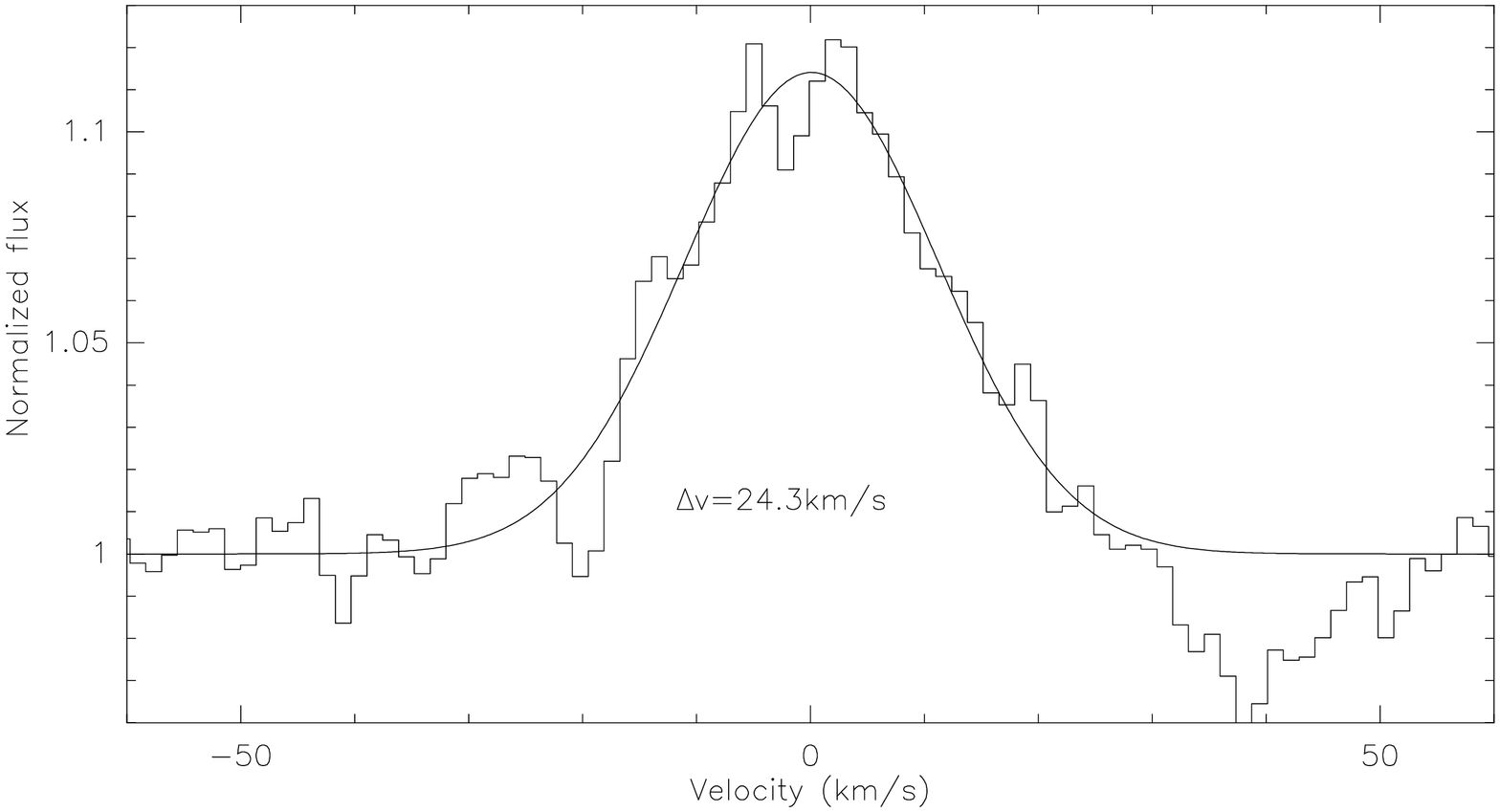}
\caption{Hydrogen recombination Pfund\,$\beta$ line of source n.}
\label{source-n_pfund}
\end{figure}

The Pfund\,$\beta$ line is also detected toward source n
(Fig.~\ref{source-n_pfund}), and we can fit a Gaussian to the
recombination line with FWHM of $\Delta v\approx 24.3$\,km\,s$^{-1}$,
a width down to 0 intensity of $\approx 70$\,km\,s$^{-1}$ (from $\sim
-35$ to $\sim +35$\,km\,s$^{-1}$) and a central velocity of
$\sim$0\,km\,s$^{-1}$.  The line is covered by 2 CRIRES chips, and
while the general Pfund\,$\beta$ line shape is the same for both
chips, the central dip at $\sim$0\,km\,s$^{-1}$ cannot independently
be reproduced. Therefore, the dip is likely only an artifact due to
insufficient signal to noise. 

The thermal line width of a $10^4$\,K hydrogen gas is
$\sim$21.4\,km\,s$^{-1}$. Convolving that with the 6\,km\,s$^{-1}$
spectral resolution, the observable thermal line width should be
$\sim$22.2\,km\,s$^{-1}$. Therefore, the measured line width does not
exceed much the thermal line width of an H{\sc ii} region. Hence, the
Pfund\,$\beta$ emission from source n does not exhibit strong
signatures from a wind but is rather consistent with a thermal H{\sc
  ii} region. We also cannot exclude that the Pfund\,$\beta$ emission
toward source n is contaminated by more broadly distributed Orion
nebula emission.

\subsection{IRC3}

\begin{figure*}
\centering
\includegraphics[width=0.98\textwidth]{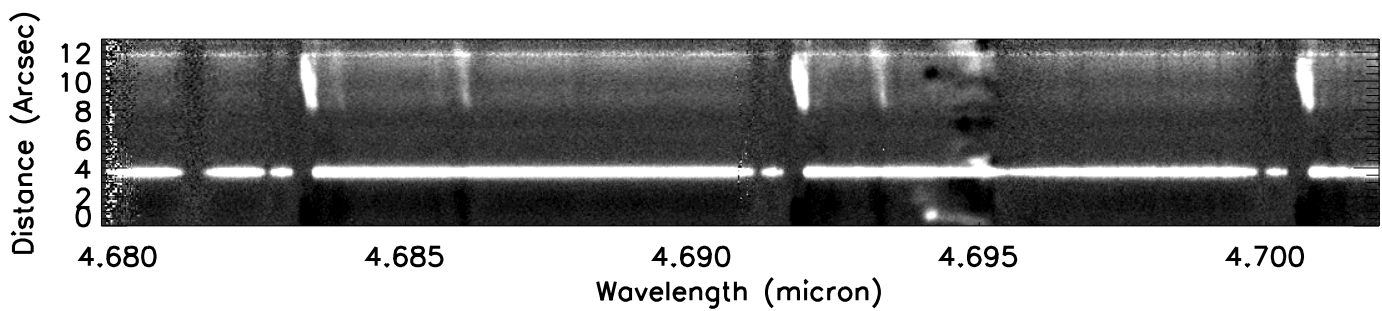}
\caption{Original data from the slit covering source n and IRC3.
  Source n is the bright continuum source at offset $\sim 4''$ whereas
  IRC3 exhibits the extended CO emission around offset $\sim 10''$.}
\label{irc3_chip}
\end{figure*}

As outlined in the Introduction, in contrast to the BN~object and
source n, the source IRC3 is unlikely to be a YSO, and it is not
prominent in any typical hot core tracer (e.g.,
\citealt{blake1996,beuther2005a}). The source rather resembles a more
extended dust density enhancement that reprocesses light from other
sources, potentially IRC2. Figure \ref{irc3_chip} presents a part of
the 2D slit spectrum clearly showing the extended nature of the CO
emission toward IRC3 in contrast to the point-like continuum structure
from source n. The continuum emission from IRC3 is very weak compared
to that from source n.

\begin{figure}
\centering
\includegraphics[width=0.36\textwidth,angle=-90]{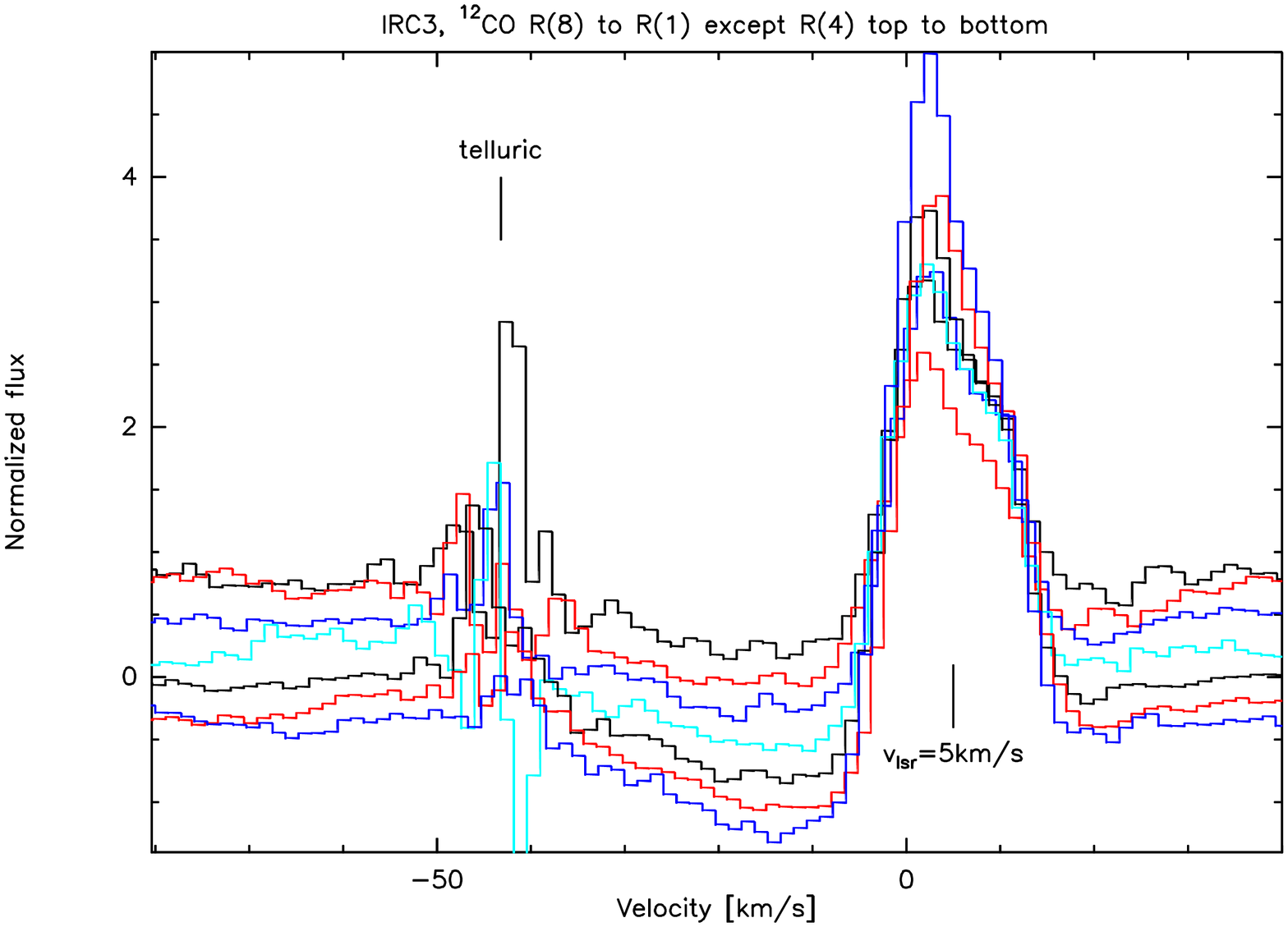}
\caption{The $^{12}$CO R(8) to R(1) lines (except of the R(4) line
  from top to bottom toward IRC3. Since the emission is extended, this
  spectrum is an average over $5.16''$ along the slit. Telluric and
  $v_{\rm{lsr}}$ velocities are marked.}
\label{irc3_rseries}
\end{figure}

\begin{figure*}
\centering
\includegraphics[width=0.99\textwidth]{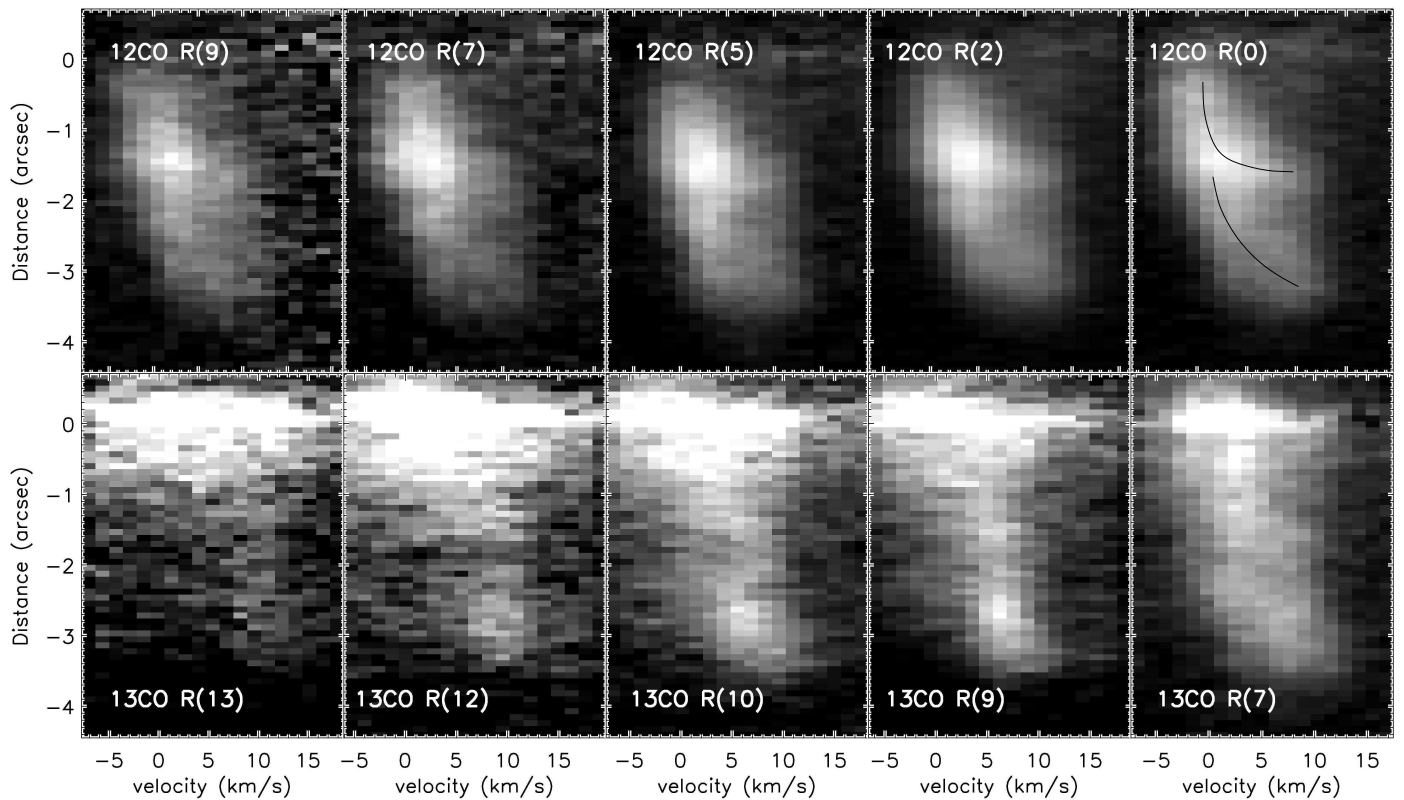}
\caption{Position velocity cuts along the slit axis for IRC3 with a PA
  of 110 degrees east of north. The top row shows diagrams for
  selected $^{12}$CO lines, and the bottom row presents the $^{13}$CO
  data. For $^{13}$CO the scaling is adapted to highlight the weaker
  extended emission in contrast to the continuum at offset $0''$. The
  wedge-like outflow structures are sketched in the top-right panel.}
\label{irc3_pv}
\end{figure*}

Figure \ref{irc3_rseries} shows the CO lines extracted as an average
spectrum of length $5.16''$ toward IRC3. While we again see a broad
absorption feature against the weaker background extending from $\sim
-5$\,km\,s$^{-1}$ out to the telluric contamination at about
$-40$\,km\,s$^{-1}$, IRC3 shows a broad and strong emission feature
peaking at $\sim 2$\,km\,s$^{-1}$. Since this emission is extended
over several arcseconds, Figure \ref{irc3_pv} presents the position
velocity diagrams of selected $^{12}$CO and all $^{13}$CO lines. In
contrast to the BN~object where the extended CO emission is just
around the rest velocity of the star (Fig.~\ref{bn_pv}), here we see a
clear trend of increasing velocity with increasing distance from the
continuum peak (the so-called Hubble law of outflows). Furthermore, in
particular the $^{12}$CO position velocity diagrams show a twofold
structure with one velocity increase to values $>10$\,km\,s$^{-1}$ at
offsets of $\sim -1.8''$ and another increase to similar velocities at
$\sim -3.3''$ offset from the main continuum peak. To guide the eye,
these wedge-like structures are sketched in the top-right panel of
Figure \ref{irc3_pv}). Such multiple wedge position velocity
structures are what jet-bow-shock entrainment models for molecular
outflows predict (e.g., \citealt{arce2006}). Furthermore, similar to
source n, the CO absorption shows also for IRC3 the extended wing-like
features toward the most blue-shifted absorption. Similar to the
pv-diagrams, where we see acceleration of the gas with distance from
the source, these wing-like absorption can also be interpreted in the
framework of accelerated winds (see also section \ref{n_co} or
\citealt{lamers1999}).  In summary, IRC3 is not only a dust density
enhancement scattering the light from another source, it may also be
part of the famous outflow emanating from the Orion-KL region.  Based
on these data, we cannot clearly define the driving source of the CO
outflow feature. While the continuum peak of IRC3 could be driving an
outflow, the CO emission may also be part of the large-scale outflow
where different driving sources have been proposed for in the
literature, e.g., source I, SMA1, or the disintegration of a bound
system once containing sources I, n and the BN-object (e.g.,
\citealt{menten1995,gomez2005,beuther2008f,zapata2009}). While the
position velocity structure of the $^{13}$CO lines largely resembles
that of the $^{12}$CO lines (Fig.~\ref{irc3_pv}), it is interesting to
note that the strongest $^{13}$CO feature is at higher velocities and
relatively distant from the continuum peak (offset $\sim -2.8''$ and
velocity $\sim +7$\,km\,s$^{-1}$. This difference is likely caused by
the varying optical depth of the $^{12}$CO and $^{13}$CO transitions.
Since the velocity of the emission feature is very close to the
$v_{\rm{lsr}}$ of the cloud (typically between 5 and 9\,km\,s$^{-1}$,
e.g., \citealt{comito2005}), self-absorption of the most abundant
$^{12}$CO isotopologue may veil this feature, whereas the less
abundant $^{13}$CO can be utilised to penetrate more deeply into the
cloud and hence detect the emission better.

\section{Discussion and conclusion}
\label{general}

High spectral resolution mid-infrared observations allow us to infer
several important characteristics of some of the major sources within
the Orion-KL region. The fundamental CO lines are largely dominated by
absorption from the individual YSO envelopes and their surrounding gas
cloud. These absorption features are blue- and red-shifted with
respect to the $v_{\rm{lsr}}$ of the molecular cloud indicating that
outflowing and inflowing gas are simultaneously present toward our
target sources.  However, we also identify interesting emission
features. Together they confirm the youth of the sources where infall
and likely accretion are still ongoing.

\paragraph{BN~object:} For the BN~object, our data confirm, at double
the spectral resolution, several of the assessments conducted already
by \citet{scoville1983}. However, also discrepancies arise.  For
example, an absorption feature reported previously around
+30\,km\,s$^{-1}$ is not found in the new data, indicating either
transient components or poor signal-to-noise in the older data.  From
a Boltzmann analysis, the rotational temperature is around 112\,K, and
we derive CO column densities of several times $10^{18}$\,cm$^{-2}$,
well in agreement with the older results based on $^{12}$CO\,v=2-0 by
\citet{scoville1983}. Using standard CO-to-H$_2$ conversion factors,
these column densities are about an order of magnitude below H$_2$
column density estimates based on mm continuum emission. As discussed
in section \ref{bn_co}, while systematics may account for some of this
discrepancy, the main difference is that the mm continuum emission is
sensitive to cold and warm dust, whereas the near-infrared absorption
lines trace only the warm gas components.

We also identify extended CO emission, likely emanating from the close
environment of BN. While the absorption stems from warm gas with
temperatures of the order 100\,K, getting the ro-vibrationally lines
in emission implies already much higher temperatures for the
corresponding gas components. The velocity range of the CO emission
between 15 and 30\,km\,s$^{-1}$ encompasses the previously inferred
velocity of the BN~object of $\sim$21\,km\,s$^{-1}$. Following
\citet{scoville1983}, who covered more J-transitions (fundamental and
overtone emission) than we do, the rotational temperature of the
compact emitting gas is around 600\,K, significantly higher than the
above derived rotational temperature of the absorbing envelope gas. As
outlined in section \ref{bn_co}, high critical densities are required
to produce this line emission by pure thermal collisional excitation,
and other processes like UV fluorescence or resonance scattering from
infrared emission may also contribute to the emission lines.
\citet{scoville1983} infer an approximate size for that emission of
$\sim$20\,AU, whereas we now resolve the emission extending to $\sim
\pm 4''$, corresponding to a diameter of $\sim 3300$\,AU. This
discrepancy can be explained by the large dynamical range between the
compact emission close to the source itself and the comparably very
weak more extended emission.  The small size found by
\citet{scoville1983} can be attributed to the strong compact emission
which also emits in the overtone bands, whereas our more sensitive new
data also detect the weaker extended features due to the high dynamic
range available with CRIRES. While the absolute ratio of the peak
emission (CO plus continuum) to the extended CO emission is $\sim$210,
the ratio of the continuum subtracted CO emission toward BN compared
to the extended CO is still 70.  \citet{scoville1983} were not
sensitive enough to detect this faint extended emission compared to
the strong central source.  Although at the spectral resolution limit
of our observations, the measured line width increase of the $^{13}$CO
emission feature with increasing excitation temperature is consistent
with a disk origin of the CO emission.  Therefore, these parameters
make the CO emitting gas potentially to be associated with a disk
around the BN~object (see also \citealt{jiang2005}).  Further
supporting the disk-outflow interpretation for the BN~object are the
Pfund\,$\beta$ hydrogen recombination line data which show broad
high-velocity line wings consistent with the decelerating outflow
scenario proposed by \citet{scoville1983}.

\paragraph{Source n:} While bright sources like the BN~object were
already feasible to be observed a while ago (e.g.,
\citealt{scoville1983}) weaker sources like source n or IRC3 allowed
reasonable high-spectral-resolution spectroscopy only with the advent
of recent instruments like CRIRES on the VLT. The general picture for
source n is relatively similar. A single broad absorption feature
extending approximately to $-65$\,km\,s$^{-1}$ traces mainly the
molecular outflow whereas red-shifted emission likely stems from an
inner infalling and accreting envelope/disk. This picture is
consistent with disk/outflow proposals for this source deduced from cm
and mid-infrared wavelength imaging projects (e.g.,
\citealt{menten1995,greenhill2004,shuping2004}). The clearly resolved
double-peaked $^{13}$CO structure allows to conduct the Boltzmann
analysis for both components. We find that the colder component has
higher H$_2$ column densities ($\sim 103$\,K \& $\sim 8\times
10^{22}$\,cm$^{-2}$) compared with the second warmer component ($\sim
163$\,K \& $\sim 5\times 10^{22}$\,cm$^{-2}$). As discussed above for
the BN~object, differences between the infrared CO derived column
densities and those estimated from mm wavelengths observations may
arise because both tracers are sensitive to gas (and dust) at
different temperatures. In contrast to the BN object, the
Pfund\,$\beta$ emission from source n is consistent with a thermal
H{\sc ii} region without a strong wind component to the line shape.

\paragraph{IRC3:} The observational signatures from the dust density
enhancement IRC3 are very different compared to the two previously
discussed sources. The continuum from IRC3 is much weaker,
nevertheless we detect CO absorption between $\sim -5$ and $\sim
-40$\,km\,s$^{-1}$. However, more importantly, we clearly detect
extended CO and $^{13}$CO emission over scales of $\sim 4''$. This
extended CO emission shows a multiple wedge-like velocity structure
consistent with jet-entrainment models of molecular outflows. Hence
IRC3 may be not merely a dust density enhancement, but it may be part
of the famous outflow from the Orion-KL region.

\paragraph{Limitations and future:} One shortcoming of the data is
that for our primary targets, the BN~object and source n, we could not
spatially resolve the inner region of the emission as done in the
lower-mass case presented in \citet{goto2006}.  The reasons may be
different for the two sources. Source n is probably still too young
and too deeply embedded so that the envelope overwhelms any emission
from the embedded disk itself. The BN object has the advantage that
its velocity of rest is offset from that of the cloud by more than
10\,km\,$^{-1}$, and hence absorption could be less of a problem for
such kind of source.  However, BN is likely significantly more
evolved, and the continuum-to-line ratio is so high that we cannot
reasonably filter out the continuum emission. Hence, we cannot well
study the inner region of the proposed disk. Nevertheless, observing
higher J-transitions as well as the CO overtone emission with todays
higher sensitivity compared to the \citet{scoville1983} observations
will likely constrain the proposed disk structure in more detail.

How to proceed now if one wants to do similar-type fundamental CO line
studies of disks in high-mass star formation? On the one hand, it is
important to not select too young sources because their envelopes will
likely almost always ``destroy'' the emission signatures. There may
exist exceptions where one views straight through the outflow cavity
face-on toward the disk. On the other hand, for more evolved regions,
the continuum emission can be very strong or maybe the remaining disk
size can be reduced again making the spatial resolution a problem.
Therefore, in addition to very careful target selections, adding AO to
achieve the best spatial resolution will be a crucial element for such
kind of studies in the coming years. Furthermore, in particular for
the mostly saturated $^{12}$CO lines, it will be important to extend
the spectral coverage to also observe higher excited CO lines, that
will likely not saturate anymore, as well as CO overtone emission.
This will allow us to better assess the hotter gas components and
hence to conduct a more detailed analysis of the $^{12}$CO data
themselves. On longer time-scales, the ELT with its proposed
mid-infrared instrument METIS promises orders of magnitude progress in
this field based on its superior sensitivity and spatial resolution.
With this instrument, we will be truly capable to resolve the gas
signatures of accretion disks around (high-mass) YSOs.

\begin{acknowledgements} 
  We like to thank a lot the anonymous referee as well as the Editor
  Malcolm Walmsley for thorough reviews which helped improving the
  paper.  H.B.~acknowledges financial support by the
  Emmy-Noether-Program of the Deutsche Forschungsgemeinschaft (DFG,
  grant BE2578).
\end{acknowledgements}

%\bibliography{/home/beuther/tex/bibliography}   
%\bibliography{/Users/henrikbeuther/paper/bibliography}
%\bibliographystyle{aa}    % this does the style, aa.bst necessary

\end{document}